\documentclass[11pt, draftcls, journal, oneside, onecolumn]{IEEEtran}
\usepackage{cite}
\ifCLASSINFOpdf
  \usepackage[pdftex]{graphicx}
\else
  \usepackage[dvips]{graphicx}
\fi
\usepackage[cmex10]{amsmath}
\usepackage{amsfonts,subfigure}
\usepackage{VarDefinitions-known-var,color}

\begin{document}

\title{Cram\'er-Rao Bound for Localization with A Priori Knowledge on Biased Range Measurements}

\author{Tao Wang
\thanks{
T. Wang is with
Communications and Remote Sensing Laboratory,
Institute of Information and Communication Technologies, Electronics and Applied Mathematics,
Universit\'e Catholique de Louvain,
1348 Louvain-la-Neuve, Belgium. E-mail: tao.wang@uclouvain.be.}
}

\markboth{Accepted by IEEE Transactions on Aerospace and Electronic Systems on 22/Dec/2010}
{}

\maketitle

\begin{abstract}
This paper derives a general expression for the Cram\'er-Rao bound (CRB) of
wireless localization algorithms using range measurements subject to bias corruption.
Specifically, the a priori knowledge about which range measurements are biased,
and the probability density functions (PDF) of the biases are assumed to be available.
For each range measurement, the error due to estimating the time-of-arrival of
the detected signal is modeled as a Gaussian distributed random variable with
zero mean and known variance.
In general, the derived CRB expression can be evaluated numerically.
An approximate CRB expression is also derived when the bias PDF is very informative.
Using these CRB expressions, we study the impact of the bias distribution on
the mean square error (MSE) bound corresponding to the CRB.
The analysis is corroborated by numerical experiments.
\end{abstract}


\IEEEpeerreviewmaketitle

\section{Introduction}

Wireless localization systems have been attracting intensive research interest
in both academia and industry lately.
For indoor or dense urban environments, Global Positioning Systems
do not function well for geolocation purposes.
Instead, a few beacons at fixed and known positions are exploited
for geolocation using measurements related to signal power decay,
bearing, difference of range, and range between a target and the beacons \cite{Bulusu00}.
Thanks to the superior multipath resolution
and penetration capability of ultra-wideband (UWB),
a range measurement can be obtained by UWB pulsed signals with high accuracy \cite{Lee02}.
Therefore, range based localization has become
a promising option for short-range wireless networks.

Much work on ranging algorithms based on time of arrival (TOA) measurements has been reported \cite{Lee02,Falsi06}.
Specifically, they are usually based on detecting the first arriving signal from a transmitter
and estimating its time of arrival (TOA) at a receiver.
Note that the clocks of the transmitter and the receiver need to synchronized,
and the transmission time at the transmitter should be sent to the receiver for
producing the range measurements.
In this paper, we consider the geolocation algorithms using range measurements
generated between the target and multiple beacons in the aforementioned way.
It is important to note that the considered algorithms are suitable for cooperative (blue-force) geolocation applications,
in that they require the cooperation of the target and the beacons in order to
synchronize their clocks for generating range measurements.

In general, a range measurement error can be modeled as the sum of two terms.
The first one is due to the error of estimating the TOA of the detected signal,
while the second one corresponds to the difference between the path length traveled
by the detected signal and the transmitter-receiver distance.
For instance, the second term is nonzero when the detected
signal does not come from the line of sight propagation path.
The first term is usually modeled as a zero mean random variable
with a Gaussian distribution in the literature \cite{Gusta05,Qi06,Jourdan08}.
As a consequence, the second term is equal to the bias of the range measurement.

It is well known that the Cram\'er-Rao bound (CRB) sets a
fundamental lower limit to the covariance of any unbiased
estimator for a vector parameter
(please refer to pages $63$ to $72$ in \cite{VanTrees}).
Specifically, the mean square error (MSE) of any unbiased location estimator is lower
bounded by the corresponding CRB's trace, referred to as
the MSE bound hereafter. The CRB and the MSE bound are used
extensively to evaluate particular localization algorithms, as
well as guide the geolocation system design fulfilling certain
accuracy requirements \cite{Gusta05,Gezici05,Wang09}.
When some of the range measurements are biased,
the CRB has been derived in \cite{Qi02NonPri} by regarding
each bias as a deterministic nuisance parameter,
and it was shown in \cite{Qi06} that using those bias-corrupted
measurements does not improve the CRB.

Recently, much work has been reported on modeling the bias as a random
variable with a particular probability density function (PDF).
Specifically, this PDF is obtained empirically by extensive experiments \cite{Jourdan05,Alavi06},
or theoretically by scattering models \cite{AlJazzar07}.
In case the PDF of the bias is known a priori,
a Cram\'er-Rao-like bound has been derived in \cite{Qi02Pri}
for the joint estimation of the deterministic target location and the random bias.
This bound is referred to as the hybrid CRB in \cite{Gini00,Reuven97},
since the parameters to be estimated consist of both deterministic and random parameters.
It was shown in \cite{Qi06} that using the bias-corrupted
measurements improves the MSE bound associated with the hybrid CRB.

Specifically, the hybrid CRB was derived from the joint
PDF of the distance measurements and the random bias.
Compared with the hybrid CRB, the CRB derived from the marginal PDF
of range measurements is much tighter and can be asymptotically attained
by a maximum likelihood (ML) estimator \cite{Reuven97,VanTrees}.
When the bias PDF is approximated by a piecewise constant function,
an expression that can be numerically evaluated was derived for the CRB in \cite{Jourdan08}.
Numerical results showed that for uniformly distributed
bias, the presence of the bias degrades the MSE bound and using
the bias-corrupted measurements improves the MSE bound.

Compared with the above existing work, this paper contains the following contributions:
\begin{itemize}
\item
We derive a general expression that can be numerically evaluated for the CRB.
When the prior bias PDF is very informative, an approximate CRB expression is derived.
\item
Based on these expressions, the impact of the bias distribution on the MSE bound
corresponding to the trace of the CRB is studied analytically.
\end{itemize}

The rest of this paper is organized as follows.
Section II describes a typical 2D localization system and related models for range measurements.
In Section III, a general CRB expression is derived for this system.
After that, an approximate CRB expression is derived in Section IV
when the prior bias PDF is very informative.
Based on these expressions,
the impact of the bias distribution on the MSE bound
is studied analytically in Section V.
We will show some numerical results in Section
VI, and complete this paper by some conclusions in Section VII.

\textit{Notations}: Upper (lower) boldface letters denote matrices (column vectors),
and $[\cdot]^T$ represents the transpose operator.
$\Grad{\mathbf{x}}{(f(\mathbf{x}))}$ stands for a column vector
which is the gradient of the function $f(\mathbf{x})$ with respect to $\mathbf{x}$,
and $\Hess{\mathbf{x}}{(f(\mathbf{x}))}$ represents a matrix
which is the Hessian of the function $f(\mathbf{x})$ with respect to $\mathbf{x}$.

\section{system setup and ranging models}

We consider a typical 2D geolocation system equipped with $\BeacNum$ beacons $(\BeacNum\geq3)$.
In the following sections, although only the CRB for this 2D system is derived,
the CRB for a 3D system can be derived in the same way and will be given as well.
The coordinate of beacon $m$ is $\Bpm = [x_m,y_m]^T (m = 1,\cdots,M)$,
and a target is located at $\Sp = [x_\mathrm{u},y_\mathrm{u}]^T$.

Suppose independent range measurements have been produced
between a target and beacons,
and the one between beacon $m$ and the target is denoted by $\rangm$.
Without loss of generality, we assume the first $\NLnum$
range measurements are biased,
while the other measurements are all unbiased.
The measurement error related to $\rangm$ is modeled as:
\begin{align}
    \sumerrm=\rangm-\dm= \left\{\begin{array}{cc}
                              \errm+\bm &  m = 1,\cdots, L \\
                              \errm     &  m = L+1, \cdots, M,
                            \end{array}\right.
\end{align}
where $\dm$ represents the distance between the target and beacon $m$,
and $\errm$ is due to the TOA estimation error of the detected signal,
and $\bm$ corresponds to the difference of the path length traveled by the detected signal
and the distance from beacon $m$ to the target.
We assume $\bm$ has the a priori known PDF $\pdfbm$
with its mean and variance denoted by $\bavgm$ and $\bstdm^2$, respectively.
In addition, we assume $\bm$ is independent of $\errm$,
and $\errm$ is Gaussian distributed with zero mean.
Besides, we denote its variance as $\stdm^2$ and assume it is known a priori.
The motivation behind this assumption is twofold.
One is that it simplifies the mathematical derivation.
The other is that the impact of the bias distribution on the localization CRB can be studied
when the TOA estimation has a guaranteed accuracy at a prescribed level.

We stack all range measurements into the column vector
$\allr=[r_1,\cdots,r_M]^T$.
The log-likelihood function of $\Sp$ is denoted as
$\logallr=\ln(\pdfallr)$ where $\pdfallr$ is the PDF of $\allr$.
Since all range measurements are independent of each other,
$\logallr=\sum_{m=1}^M\logrm$,
where $\logrm=\ln(\pdfrm)$
is the log-likelihood function of $\rangm$ given $\Sp$,
and $\pdfrm$ is the PDF of $\rangm$.

For an unbiased range measurement $\rangm\;(m=\NLnum+1,\cdots,M)$,
$\pdfrm$ can be formulated as:
\begin{equation}
\pdfrm=\frac{1}{\sqrt{2\pi\stdm^2}}\exp\left[-\frac{(\rangm-\dm)^2}{2\stdm^2}\right]  \end{equation}

For a biased range measurement $\rangm\;(m=1,\cdots,\NLnum)$,
we define $\logrbm=\ln(\pdfrbm)$
as the joint log-likelihood function of $\rangm$ given $\Sp$ and $\bm$,
where $\pdfrbm$ is the PDF of $\rangm$ conditional on $\bm$ and $\Sp$.
$\pdfrbm$ and $\pdfrm$ can be expressed respectively as:
\begin{eqnarray}
\pdfrbm&=&\frac{1}{\sqrt{2\pi\stdm^2}}\exp\left[-\frac{(\rangm-\dm-\bm)^2}{2\stdm^2}\right]  \\
\pdfrm&=&\intinf{\pdfrbm\pdfbm}{\bm}          \label{eq:pdfrm}
\end{eqnarray}

\section{Derivations of a general crb expression}\label{sec:CRB-derivation}

The CRB is the inverse of the Fisher information matrix (FIM) $\FIM$,
which is expressed by:
\begin{equation}
    \FIM=\AVER{\allr;\Sp}{(\Grad{\Sp}{\logallr})(\Grad{\Sp}{\logallr})^T}          \label{eq:FIM}
\end{equation}
\noindent
where $\AVER{\allr;\Sp}{\cdot}$ denotes the expectation operator with respect to $\pdfallr$.
Since all distance measurements are independent of each other,
$\FIM=\sum_{m=1}^M\FIMm$ where $\FIMm$ is the FIM related to $\rangm$ \cite{VanTrees}:
\begin{equation}
    \FIMm=\AVER{\rangm;\Sp}{(\Grad{\Sp}{\logrm})(\Grad{\Sp}{\logrm})^T}          \label{eq:FIMm}
\end{equation}
\noindent
where $\AVER{\rangm;\Sp}{\cdot}$ denotes the expectation operator with respect to $\pdfrm$.

When $m = \NLnum+1,\cdots,\BeacNum$, we can show that:
\begin{eqnarray}
\FIMm&=&\AVER{\rangm;\Sp}{\frac{(\rangm-\dm)^2}{\stdm^4}\Orim\Orim^T}  \nonumber\\
     &=&\stdm^{-2}\Orim\Orim^T,
\end{eqnarray}
where $\Orim = \frac{\Sp-\Bpm}{\dm}$ represents the orientation of the target relative to beacon $m$.

When $m = 1,\cdots,\NLnum$, we can first prove that:
\begin{eqnarray}
\Grad{\Sp}{\logrm}&=&\frac{1}{\pdfrm}\intinf{\left[\Grad{\Sp}{\pdfrbm}\right]\pdfbm}{\bm}                 \nonumber\\
                   &=&\intinf{\frac{\pdfrbm\pdfbm}{\pdfrm}\left[\Grad{\Sp}{\logrbm}\right]}{\bm}           \nonumber\\
                   &=&\AVER{\bm|\rangm;\Sp}{\Grad{\Sp}{\logrbm}}                     \nonumber\\
                   &=&\stdm^{-2}\posterrm\Orim             \label{eq:Grad-NL}
\end{eqnarray}
\noindent
where $\posterrm=\AVER{\bm|\rangm;\Sp}{\rangm-\dm-\bm}$ and
$\AVER{\bm|\rangm;\Sp}{\cdot}$ denotes the expectation operator with respect to
the posterior PDF $\pdfbrm$.
This means that $\posterrm$ actually stands for the posterior mean
of $\errm$ conditional on $\rangm$.
Inserting \eqref{eq:Grad-NL} into \eqref{eq:FIMm},
we can compute $\FIMm$ by:
\begin{equation}
\FIMm=\coeffm\Orim\Orim^T      \label{eq:FIMm-matrix}
\end{equation}
\noindent
where $\coeffm=\stdm^{-4}\cdot\AVER{\rangm;\Sp}{\posterrm^2}$.

Summing $\FIMbegin, \cdots, \FIMend$ and taking the inverse,
we can express the CRB as:
\begin{equation}
\CRB=\inverse{\sum_{m=1}^\NLnum\coeffm\Orim\Orim^T + \sum_{m=\NLnum+1}^\BeacNum\stdm^{-2}\Orim\Orim^T}    \label{eq:CRB}
\end{equation}
\noindent

The CRB for a $3$-D localization system can be derived in the same way as given above.
Most interestingly, we can show after simple mathematical arrangement that
the CRB for the $3$-D localization system can still be evaluated by \eqref{eq:CRB},
when $\Sp$ and $\Bpm$ are substituted with the $3$-D coordinates of the target
and beacon $m$, respectively.

It is important to note that \eqref{eq:CRB} is a general expression for the CRB,
when each biased measurement $\rangm$ is Gaussian distributed conditional on $\bm$.
However, we have proven that
$\Grad{\Sp}{\logrm}=\AVER{\bm|\rangm;\Sp}{\Grad{\Sp}{\logrbm}}$
holds in general for any particular form of $\pdfrbm$.
This means that a closed-form CRB expression can be derived in a similar way
when $\pdfrbm$ takes other specific form.

To illustrate the generality of \eqref{eq:CRB}, we show that
the CRB can be derived by \eqref{eq:CRB} when $\bm$ is deterministic.
Note that this CRB will be used when illustrating the effectiveness of the approximate
CRB derived in Section IV.
In this case, $\pdfbrm=\delta(\bm-\bavgm)$, $\posterrm=\rangm-\dm-\bavgm$, $\pdfrm=p(\rangm|\bm=\bavgm;\Sp)$,
and therefore $\coeffm=\stdm^{-2}$.
This means that the contribution of $\rangm$ to the CRB is the same as if
this range measurement were not bias corrupted.
This is true since when $\bm$ is deterministic and known a priori, its value can be
subtracted from $\rangm$, which means that $\rangm$ is in effect not subject to the bias corruption.

In general, an analytical expression for $\coeffm$ does not exist,
in such a case $\coeffm$ can be evaluated numerically, with $\CRB$ following from equation (10).
This provides a way to study the effects of the geometric configuration,
the TOA estimation quality, as well as the bias distribution on the CRB.
To this end, it can easily be shown that
\begin{align}
\coeffm=\stdm^{-4}\intinf{\posterrm^2\pdfrm}{\rangm}           \label{eq:coeffm-num}
\end{align}
and
\begin{align}
\posterrm=\intinf{(\rangm-\dm-\bm)\frac{\pdfrbm\pdfbm}{\pdfrm}}{\bm},       \label{eq:posterr-num}
\end{align}
respectively. For every value of $\rangm$,
$\pdfrm$ and $\posterrm$ can be first evaluated by computing numerically the integrals
in \eqref{eq:pdfrm} and \eqref{eq:posterr-num} with respect to $\bm$
\footnote{The detailed procedure of numerical integration can be found in pages $198-200$ in \cite{NumIntegral}.}, respectively.
Then, $\coeffm$ can be evaluated by computing numerically the integral
with respect to $\rangm$ in \eqref{eq:coeffm-num}.

Note that the CRB expression in \eqref{eq:CRB}, as those presented
in \cite{Gusta05,Qi06,Jourdan08}, is derived under the assumption that
we have the a priori knowledge about which range measurements are biased.
In almost all practical scenarios, this a priori knowledge may not be available \cite{Hernandez02,Zhang05,Hernandez06}.
In this case, the CRB in \eqref{eq:CRB} is an optimistic lower bound,
since it assumes the implicit knowledge which is not really available for the range measurements.
We will illustrate the optimism of \eqref{eq:CRB} when the a priori knowledge about
which range measurements are biased is not available by numerical experiments in Section VI.
Relaxing the assumption of knowing which measurements are biased,
and generating an information reduction factor that scales the CRB
when the probability of each measurement getting biased is known a priori,
in a similar way as shown in \cite{Hernandez02,Zhang05,Hernandez06},
would be a valuable piece of future work.

We should also note that the derived CRB is for cooperative geolocation applications,
since the synchronization between the target and the beacons is required
as mentioned in the introduction part.
For geolocating evasive targets, range measurements would only be available with active sensing,
in which case azimuth measurements, etc., would be available as well.
It is another interesting piece of future work to derive the CRB for
localization using azimuth measurements, etc., in addition to range measurements.

\section{Derivation of an approximate crb expression when $\bstdm\ll\stdm$}

We now derive an approximate CRB expression when $\bm$ has a small variance
compared to $\stdm$ (i.e. $\bstdm\ll\stdm$).
First of all, we express $\FIMm (m=1,\cdots,\NLnum)$ alternatively as \cite{VanTrees}:
\begin{equation}
    \FIMm=-\AVER{\rangm;\Sp}{\Hess{\Sp}{\logrm}}          \label{eq:FIMm2}
\end{equation}

Next, we approximate $\pdfrbm$ by the first-order Taylor-series expansion as follows:
\begin{equation}
    \pdfrbm\approx\pdfrbmean+\delpdfrbmean\cdot(\bm-\bavgm)          \label{eq:Taylor}
\end{equation}
\noindent
where $\pdfrbmean$ and $\delpdfrbmean$ are expressed respectively as:
\begin{eqnarray}
\pdfrbmean&=&p(\rangm|\bm=\bavgm;\Sp)                                   \\
\delpdfrbmean&=&\frac{\partial\pdfrbm}{\partial\bm}\Big|_{\bm=\bavgm}
\end{eqnarray}

Based on \eqref{eq:pdfrm} and \eqref{eq:Taylor}, we can show that $\pdfrm\approx\pdfrbmean$.
In addition, we make the following approximations:

\setlength{\arraycolsep}{0pt}
\begin{eqnarray}
\frac{\pdfrbm}{\pdfrm}&\approx&\frac{\pdfrbmean+\delpdfrbmean\cdot(\bm-\bavgm)}{\pdfrbmean}             \nonumber \\
                          &=&1+\frac{\partial\logrbm}{\partial\bm}\Big|_{\bm=\bavgm}\cdot(\bm-\bavgm)       \nonumber \\
                          &=&1+\frac{\rangm-\dm-\bavgm}{\stdm^2}(\bm-\bavgm)           \label{eq:pdfratio}
\end{eqnarray}
\setlength{\arraycolsep}{5pt}

Inserting \eqref{eq:pdfratio} into \eqref{eq:posterr-num},
we can approximate $\posterrm$ as follows:
\begin{equation}
\posterrm\approx\left(1-\frac{\bstdm^2}{\stdm^2}\right)(\rangm-\dm-\bavgm)       \label{eq:am-approx}
\end{equation}

Based on \eqref{eq:Grad-NL} and \eqref{eq:am-approx},
$\Hess{\Sp}{\logrm}$ can be approximated by:
\begin{eqnarray}
\Hess{\Sp}{\logrm}&=&\stdm^{-2}\left[(\Grad{\Sp}{\posterrm})\Orim^T+\posterrm(\Hess{\Sp}{\dm})\right]   \nonumber\\
                  &\approx&\stdm^{-2}\left[-\left(1-\frac{\bstdm^2}{\stdm^2}\right)\Orim\Orim^T+\posterrm(\Hess{\Sp}{\dm})\right]  \label{eq:Hesslogrm}
\end{eqnarray}

Using \eqref{eq:am-approx}, \eqref{eq:Hesslogrm}, and \eqref{eq:FIMm2},
$\FIMm$ can be approximated by:
\begin{equation}
\FIMm\approx\coeffappm\Orim\Orim^T   \label{eq:FIMmappr}
\end{equation}
\noindent
where
\begin{align}
\coeffappm=\stdm^{-2}\left(1-(\bstdm/\stdm)^2\right) \label{eq:coeffamappr}.
\end{align}

From \eqref{eq:FIMm-matrix} and \eqref{eq:FIMmappr},
we can see that $\coeffm\approx\coeffappm$.
In addition, CRB can be approximated by:
\begin{equation}
\CRB\approx\inverse{\sum_{m=1}^\NLnum\coeffappm\Orim\Orim^T + \sum_{m=\NLnum+1}^\BeacNum\stdm^{-2}\Orim\Orim^T}    \label{eq:CRBappr}
\end{equation}
\noindent

Note that a better approximation of the CRB can be derived with a higher order
Taylor-series expansion in the same way as given above,
when the statistical moments of $\bm$ higher than the $2$nd order are known a priori.
Here, we only use the first order Taylor-series expansion to derive an approximation
when the $2$nd order moment of $\bm$ (i.e. $\bstdm$) is known.
The derived approximate CRB expression features a simple mathematical structure
but reveals the influence of the bias distribution on the CRB clearly.

We will illustrate the effectiveness of \eqref{eq:CRBappr} by numerical experiments
using measured bias distributions in Section V.
In fact, we can analytically illustrate the effectiveness of \eqref{eq:CRBappr}
when $\bm$ is Gaussian distributed.
Note that although in practice the Gaussian distribution is unlikely to be
a good approximation of the true bias distribution,
it lends the CRB to be easily simplified analytically so that
the effectiveness of \eqref{eq:CRBappr} can be examined.
In this case, $\rangm$ can be equivalently modeled as $\rangm = \dm + \bavgm + \errm'$
where $\errm'$ is Gaussian distributed with zero mean and variance $\stdm^2+\bstdm^2$.
Therefore, $\coeffm=(\stdm^2+\bstdm^2)^{-1}$ according to the earlier analysis
for the case when the bias is deterministic in Section III.
Since $\bstdm\ll\stdm$,
we can see that $\CRB$ can indeed be approximately computed with \eqref{eq:CRBappr},
due to the fact that
\begin{eqnarray}
\coeffm&=&\stdm^{-2}\left(1+\frac{\bstdm^2}{\stdm^2}\right)^{-1}           \nonumber\\
       &\approx&\stdm^{-2}\left(1-\frac{\bstdm^2}{\stdm^2}\right)                \nonumber\\
       &\approx&\coeffappm.
\end{eqnarray}

\section{Mse bound analysis}

It is important to note that the MSE bound, namely the trace of $\CRB$ and denoted by $\trace{\CRB}$,
reduces as $\coeffm$ increases.
This is because $\sum_{m=1}^M\FIMm$ increases in a positive definite sense,
therefore both ${\CRB}$ and $\trace{\CRB}$ decrease,
which means that the MSE bound improves, as $\coeffm$ increases.

When $\rangm\;(m=1,\cdots,\NLnum)$ are discarded,
we can compute $\CRB$ by setting $\coeffm$ to $0$.
Since $\coeffm$ is no smaller than $0$,
using a bias-corrupted measurement results in the same or better MSE bound
than discarding it.

An important property of $\coeffm$ is that $\coeffm\leq\stdm^{-2}$.
This can be justified by:
\begin{eqnarray}
\coeffm&=&\stdm^{-4}\cdot\AVER{\rangm;\Sp}{\left(\AVER{\bm|\rangm;\Sp}{\rangm-\dm-\bm}\right)^2}            \nonumber \\
       &\leq&\stdm^{-4}\cdot\AVER{\rangm;\Sp}{\AVER{\bm|\rangm;\Sp}{(\rangm-\dm-\bm)^2}}                 \nonumber \\
       &=&\stdm^{-4}\cdot\AVER{\bm}{\AVER{\rangm|\bm;\Sp}{(\rangm-\dm-\bm)^2}}                         \nonumber \\
       &=&\stdm^{-2}
\end{eqnarray}
\noindent
where $\AVER{\bm}{\cdot}$ and $\AVER{\rangm|\bm;\Sp}{\cdot}$ stand for
the expectation operator with respect to $\pdfbm$ and $\pdfrbm$, respectively.
Specifically, the second inequality is according to the Jensen's inequality,
and the equality holds if $\bm$ is deterministic (please refer to pages $77-78$ in \cite{Convex-opt}).
This implies the presence of the random bias $\bm$ degrades the MSE bound.

It is interesting to consider two special cases:
\begin{enumerate}[\IEEEsetlabelwidth{2}]
\item
The first case is when $\bm$ has a very large variance compared to $\stdm$ (i.e., $\bstdm\gg\stdm$)
so that for any fixed $\rangm$, $\pdfbm$ is approximately constant
within the nonzero support of $\pdfrbm$.
This case corresponds to the scenario where the prior bias PDF is not informative,
and we can find that $\coeffm\approx0$.
If $\bm$ is Gaussian distributed,
this is the case when $\bstdm^2$ is sufficiently large, and $\coeffm$ is indeed close to zero.
This means that discarding a bias-corrupted measurement
degrades the MSE bound slightly when the prior bias PDF is not informative.
\item
The second case is when $\bm$ has a very small variance compared to $\stdm$ (i.e., $\bstdm\ll\stdm$)
so that for any fixed $\rangm$, $\pdfrbm$ varies slowly with respect to $\bm$ within
the nonzero support of $\pdfbm$.
This case corresponds to the scenario where the prior bias PDF is very informative.
Based on the analysis in Section IV,
we can find that $\coeffm\approx\coeffappm=\stdm^{-2}\left(1-(\bstdm/\stdm)^2\right)\approx\stdm^{-2}$.
This means that the presence of the random bias $\bm$ results in a slight degradation to the MSE bound,
compared to the case in which there are no measurement biases.
\end{enumerate}

\section{Numerical experiments}\label{sec:numerical-experiment}

\begin{figure}
\centering
\includegraphics[width=0.5\textwidth]{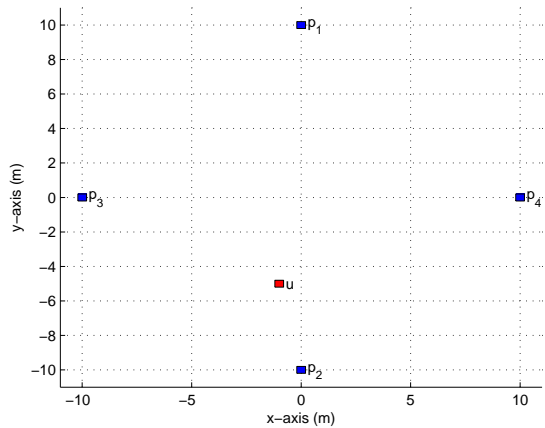}
\caption{
The localization system with $4$ beacons considered in the numerical experiments.
Note that only $r_1$, namely the range measurement between $\Sp$ and ${\bf p}_1$, is biased.}\label{fig:system}
\end{figure}

\begin{figure}
\centering
\includegraphics[width=0.5\textwidth]{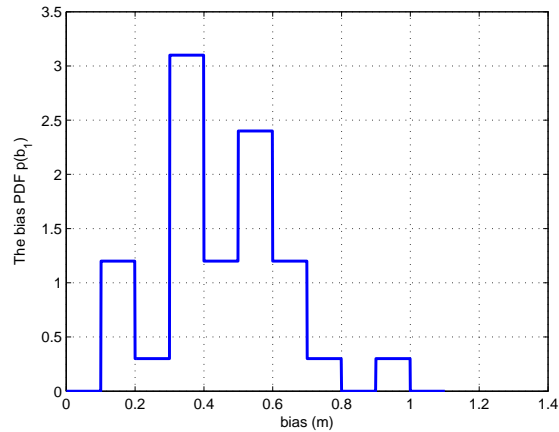}
\caption{
The PDF of $b_1$, which is the bias of the range measurement between $\Sp$ and ${\bf p}_1$,
when $\Delta=0.1$ m.}\label{fig:bias}.
\end{figure}

\begin{figure}
\centering
\includegraphics[width=0.5\textwidth]{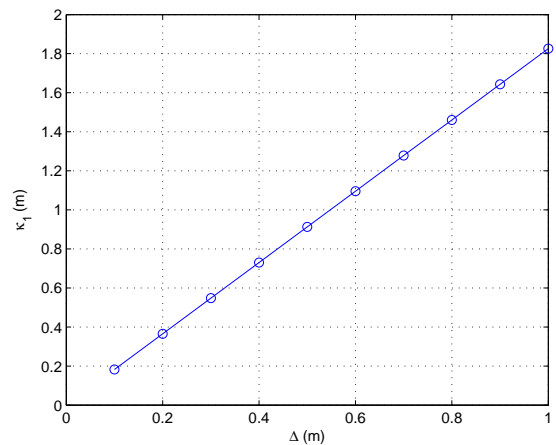}
\caption{
$\kappa_1$, which is the standard deviation of the random bias $b_1$,
when $\Delta$ increases from $0.1$ to $1$ m.}\label{fig:biasstd}
\end{figure}

For illustration purposes, we consider a wireless localization system with four beacons
located respectively at $\beacone$, $\beactwo$, $\beacthr$, $\beacfou$,
and a target at $\targetpos$, as shown in Figure \ref{fig:system}.
We assume $r_1$ is biased while $r_2$, $r_3$, and $r_4$ are unbiased, and $\stdm=1\,(m=1,\cdots,4)$.
$p(b_1)$ is assumed to have the same shape as the measured bias distribution reported in \cite{Jourdan05}.
Specifically, $p(b_1)$ is expressed as
\begin{align}
p(b_1) = \left\{ \begin{array}{cc}
                    0          &    {\rm if\;} b_1\in(-\infty,\Omega_0]\cup(\Omega_9,\infty) \\
           \frac{P_i}{\Delta}  &    {\rm if\;} b_1\in(\Omega_i, \Omega_{i+1}] {\;\rm with\;} i=0,\cdots,8,
                 \end{array}
         \right.
\end{align}
where $\Omega_i = 0.1 + i\Delta$, and $P_i$ is given in Table \ref{tab:bias} for $i=0,\cdots,8$.
When $\Delta=0.1$, $p(b_1)$ is shown in Figure \ref{fig:bias}.
It can be readily derived that $\overline{b}_1=0.1+3.49\Delta$ and $\kappa_1=1.83\Delta$,
which means that $\kappa_1$ is a linear function of $\Delta$.
When $\Delta$ increases from $0.1$ to $1$, we have computed $\kappa_1$,
and the results are shown in Figure \ref{fig:biasstd}.
Note that all the above values related to a coordinate or length have the units of meters.

\begin{table}
\centering
\caption{The value of $P_i$, $i=0,\cdots,8$.}\label{tab:bias}
\begin{tabular}{c|c|c|c|c|c|c|c|c}
  \hline
  $P_0$  & $P_1$  & $P_2$  & $P_3$  & $P_4$  & $P_5$  & $P_6$  & $P_7$ & $P_8$\\
  \hline
  $0.12$ & $0.03$ & $0.31$ & $0.12$ & $0.24$ & $0.12$ & $0.03$ & $0$   & $0.03$\\
  \hline
\end{tabular}
\end{table}

\begin{figure}
  \centering
    \subfigure[when $\kappa_1/\stdm<1$]{
     \includegraphics[width=0.5\textwidth]{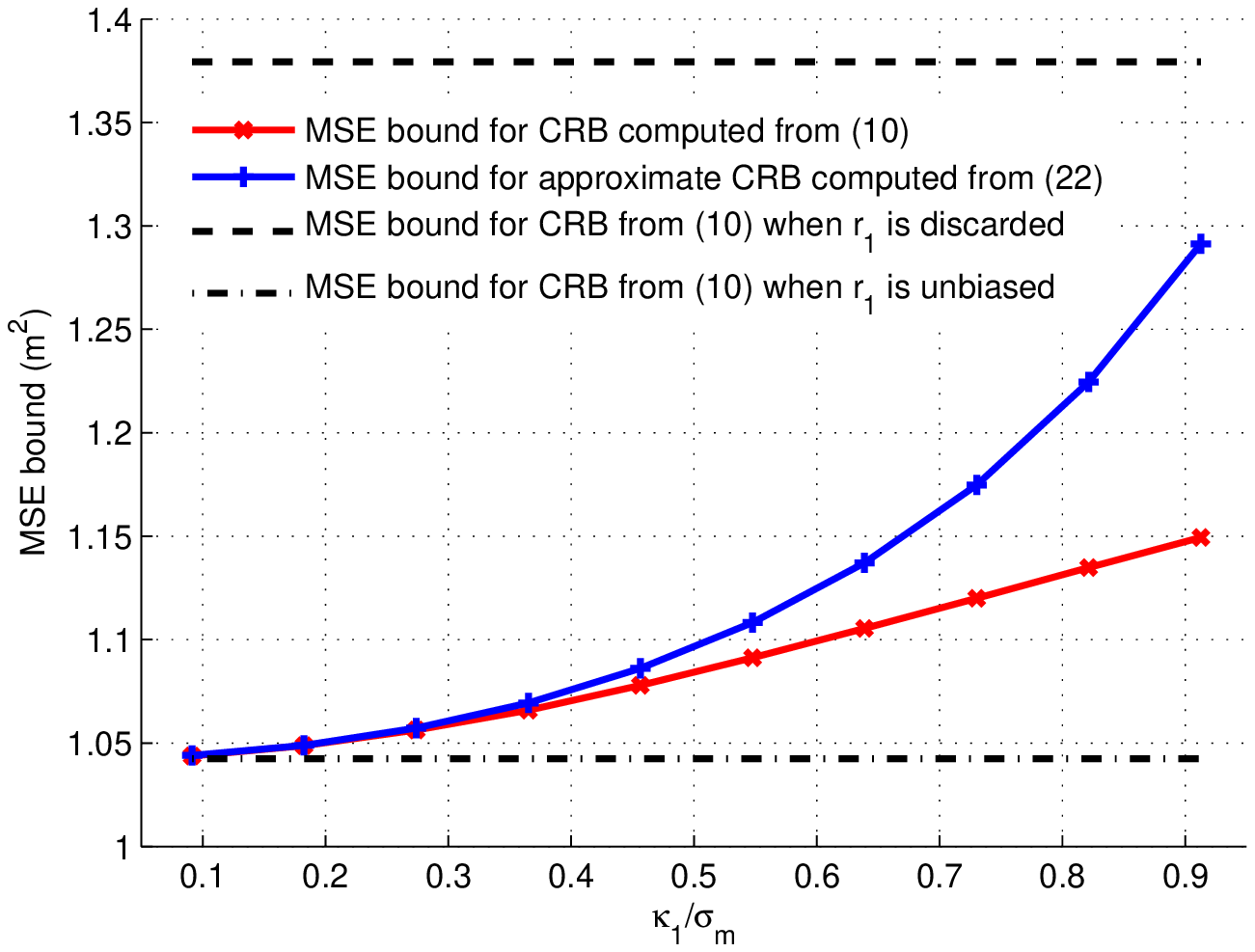}}
    \subfigure[when $\kappa_1/\stdm>1$]{
     \includegraphics[width=0.5\textwidth]{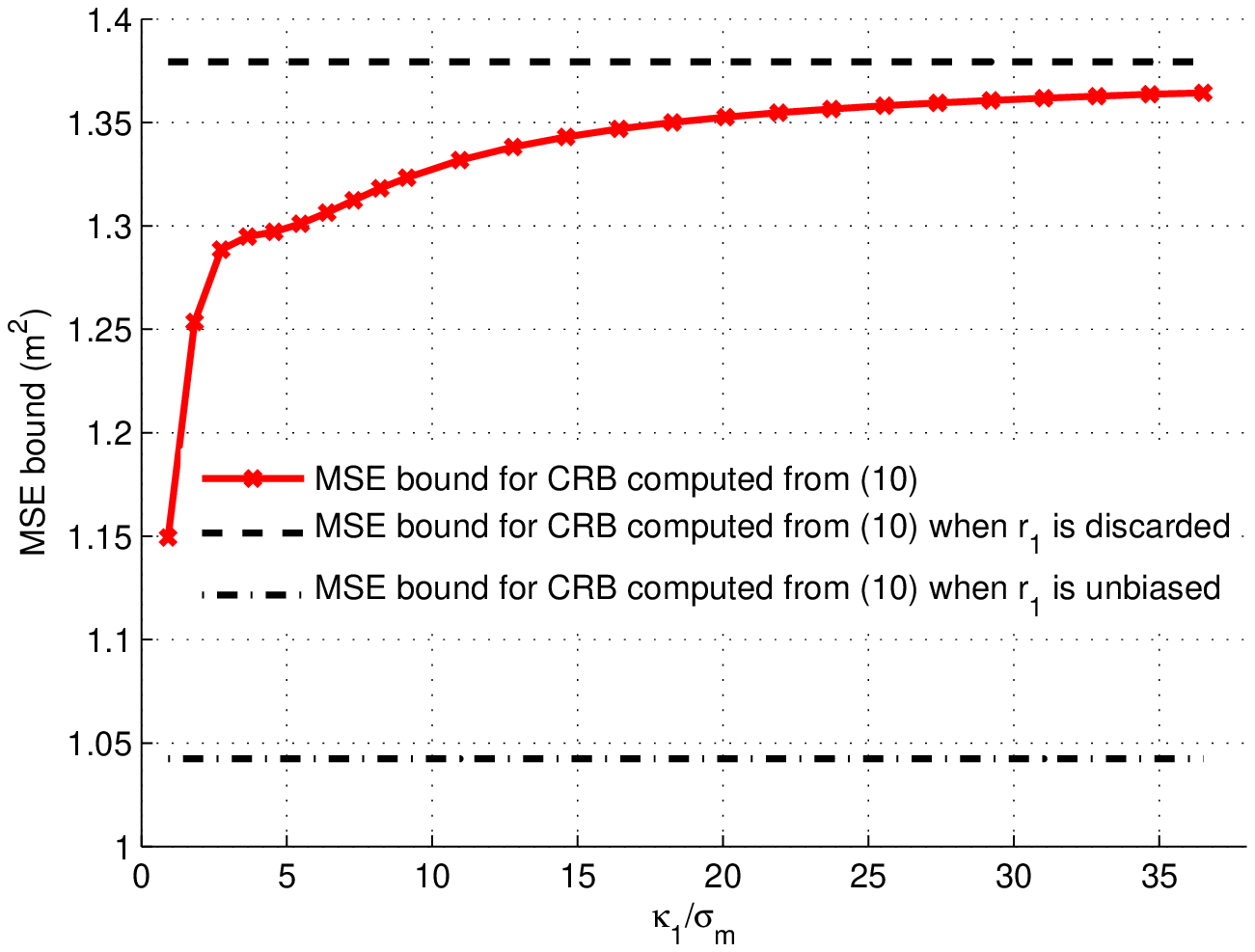}}
     \caption{The MSE bounds corresponding to the CRBs computed when $\kappa_1/\stdm<1$ and $\kappa_1/\stdm>1$, respectively. }
\label{fig:CRB}
\end{figure}

To examine the effectiveness of the approximate CRB expression \eqref{eq:CRBappr},
we have evaluated the CRB from \eqref{eq:CRB} numerically,
and the approximate CRB from \eqref{eq:CRBappr} when $\Delta$ increases with $\kappa_1/\stdm<1$ satisfied.
The MSE bounds corresponding to those CRBs are shown with respect to $\kappa_1/\stdm$ in Figure \ref{fig:CRB}.a.
We can see that when $\kappa_1/\stdm\leq0.5$, the MSE bound corresponding to the approximate CRB
is very close to the one computed from \eqref{eq:CRB}.
This illustrates the effectiveness of \eqref{eq:CRBappr} when $\kappa_1\ll\stdm$.

To examine the analysis in Section V,
we have also evaluated the CRB from \eqref{eq:CRB} numerically when $\Delta$ increases with $\kappa_1/\stdm>1$ satisfied,
and the corresponding MSE bound is shown with respect to $\kappa_1/\stdm$ in Figure \ref{fig:CRB}.b.
We have also computed the CRBs when $r_1$ is discarded and when $r_1$ is not bias corrupted, respectively,
and shown the corresponding MSEs in both Figure \ref{fig:CRB}.a and Figure \ref{fig:CRB}.b.
It is clearly shown in Figure \ref{fig:CRB}.a that
as $\kappa_1$ reduces, the MSE bound approaches the one corresponding to the case when $r_1$ is not bias corrupted.
On the other hand, as $\kappa_1$ increases, the MSE bound approaches the one corresponding to the case when $r_1$ is discarded,
as shown in Figure \ref{fig:CRB}.b.
These observations corroborate the analysis in Section V.

\begin{figure}
\centering
\includegraphics[width=0.5\textwidth]{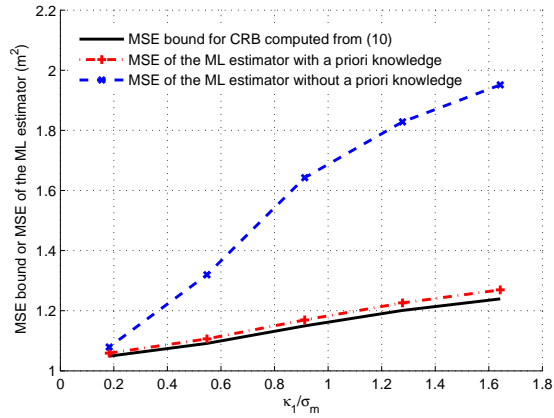}
\caption{
The MSEs of the ML location estimators and the MSE bound corresponding to the CRB computed from (10).
Note that the MSE bound is the same as the corresponding one depicted in Fig. \ref{fig:CRB}.}
\label{fig:ML}
\end{figure}

Note that the derived CRB is an optimistic bound in scenarios
when we do not have the a priori knowledge about which range measurements are biased.
To illustrate this remark explicitly, we compare the MSE bound corresponding to the CRB computed from \eqref{eq:CRB}
with the MSE of two ML location estimators as $\kappa_1/\stdm$ increases,
since the ML estimator is widely used and able to achieve the corresponding CRB asymptotically.
For the first ML estimator, the above mentioned a priori knowledge is not available,
and a binary parameter $s_m$ is introduced to indicate that $r_m$ ($m=1,\cdots,4$) is bias corrupted if $s_m=0$.
Specifically, the first ML estimator produces a joint estimate of $\Sp$ and
$\{s_m,m=1,\cdots,4\}$ as the maximizer of the log-likelihood function
\begin{align}
L(\Sp,s_1,s_2,s_3,s_4) = \sum_{m=1}^4{L_m(\Sp,s_m)},
\end{align}
where
\begin{align}
L_m(\Sp,s_m) =  \ln(\frac{1}{\sqrt{2\pi\stdm^2}}) + \ln\bigg(s_m\exp\{-\frac{(\rangm-\dm)^2}{2\stdm^2}\}        \nonumber\\
                    +  (1-s_m)\intinf{\pdfbm\exp\{-\frac{(\rangm-\dm-\bm)^2}{2\stdm^2}\}}{\bm}\bigg),
\end{align}
and $\pdfbm=p(b_1)$, $m=2,3,4$.
For the second ML estimator, the a priori knowledge is available, i.e.,
this estimator knows a priori that only $r_1$ is biased, and produces an estimate of $\Sp$ as
the maximizer of $L(\Sp,s_1,s_2,s_3,s_4)$ when $s_1=0$ and $s_m=1$, $m=2,3,4$.
For every $\Delta$, the MSE of each estimator is computed by
averaging the square errors of the location estimates for $1000$ random
realizations of range measurements, and the results are shown in Figure \ref{fig:ML} with respect to $\kappa_1/\stdm$.
We can see that the MSE of the second ML estimator is slightly above the MSE bound
computed from \eqref{eq:CRB}, while the MSE of the first ML estimator is much greater
than that MSE bound.
These observations indicate that the derived CRB is indeed an optimistic lower bound
for localization algorithms without the a priori knowledge about which range measurements are biased.

\section{Conclusion}
We have derived a general expression for the CRB of wireless localization algorithms
using range measurements subject to bias corruption.
Specifically, the knowledge about which range measurements are biased,
and the probability density functions (PDF) of the biases are assumed to be known a priori.
For each range measurement, the error due to estimating the TOA of
the detected signal is modeled as a Gaussian distributed random variable with
zero mean and known variance.
We have also derived an approximate CRB expression when the bias PDF is very informative.
Using these CRB expressions, we have studied the impact of the bias distribution on
the MSE bound corresponding to the CRB.
The analysis has been corroborated by numerical experiments.

\section*{Acknowledgement}

The author is very grateful to Prof. Claude Jauffret for coordinating the review,
as well as the anonymous reviewers for their valuable comments and suggestions
to improve the quality of this paper.

\ifCLASSOPTIONcaptionsoff
  \newpage
\fi

\bibliographystyle{IEEEtran}
\bibliography{CRBRangeBias-Reference}

\begin{IEEEbiography}[{\includegraphics[width=1in,height=1.5in,clip,keepaspectratio]{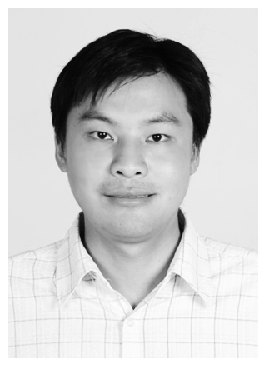}}]{Tao Wang}
received respectively B.E. and DoE degree in electronic engineering
from Zhejiang University, China, in 2001 and 2006,
as well as civil electrical engineering degree ({\it summa cum laude})
from Universit{\'e} Catholique de Louvain, Belgium.
He had multiple research appointments
in Motorola Electronics Ltd. Suzhou Branch, China, since 2000 to 2001,
the Institute for Infocomm Research, Singapore, since 2004 to 2005,
Delft University of Technology and Holst Center in the Netherlands since 2008 to 2009.
He is now a researcher in Universit\'e Catholique de Louvain, Belgium.
He has been an associate editor in chief for Signal Processing: An International Journal (SPIJ)
since Oct. 2010.
His current research interests are in the optimization of wireless localization systems
with energy awareness, as well as resource allocation algorithms in wireless communication systems.
\end{IEEEbiography}


\end{document}